\documentclass[12pt,preprint]{aastex}
\begin{document}
\title{A Multidimensional Dependence of the Substructure Evolution on the Tidal Coherence}
\author{Jounghun Lee}
\affil{Astronomy Program, Department of Physics and Astronomy, Seoul National University, 
Seoul 08826, Republic of Korea} 
\email{jounghun@astro.snu.ac.kr}
\begin{abstract}
We numerically explore how the subhalo mass-loss evolution is affected by the tidal coherences measured along different 
eigenvector directions.  The mean virial-to-accretion mass ratios of the subhalos are used to quantify the severity of their mass-loss evolutions 
within the hosts, and the tidal coherence is expressed as an array of three numbers each of which quantifies the alignment 
between the tidal fields smoothed on the scales of $2$ and $30\,h^{-1}$Mpc in each direction of three principal axes. 
Using a Rockstar halo catalog retrieved from a N-body simulation, we investigate if and how the mass-loss evolutions of the subhalos 
hosted by distinct halos at fixed mass scale of [$1$-$3$]$10^{14}\,h^{-1}\,M_{\odot}$ are correlated with three components of the tidal coherence. 
The tides coherent along different eigenvector directions are found to have different effects on the subhalo mass-loss evolution,
which cannot be ascribed to the differences in the densities and ellipticities of the local environments. 
It is shown that the substructures surrounded by the tides highly coherent along the first eigenvector direction and highly {\it incoherent} 
along the third eigenvector direction experience the least severe mass-loss evolution, while the tides highly {\it incoherent} only along the 
first eigenvector direction is responsible for the most severe mass-loss evolution of the subhalos. 
Explaining that the coherent tides have an obstructing effect on the satellite infalls onto their hosts and that the strength of the obstruction 
effect depends on which directions the tides are coherent or {\it incoherent} along, we suggest that the multidimensional dependence 
of the substructure evolution on the tidal coherence should be deeply related to the complex nature of the large-scale assembly bias.
\end{abstract}
\section{Introduction}\label{sec:intro}

The classical excursion set theory based on the standard $\Lambda$CDM (cosmological constant $\Lambda$ and cold dark matter) 
model provided an analytical framework within which the formation and evolution of DM halos, the building blocks of the large-
scale structure in the universe, can be physically tracked down \citep{PS74,bbks86,bon-etal91,BM96,SMT01}. According to this theory, 
the hierarchical accretion and merging events, which are the dominant driver of the halo growth, owe their frequencies solely to the halo masses. 
N-body simulations that were performed to complement the theory with desired accuracy and precision, however, invalidated this simple picture, 
discovering a puzzling phenomenon, so called the "halo assembly bias": The clustering strength of the DM halos affect their formation epochs 
and growth rates on the same mass scale  \citep{GW07}. 
Although the discovery of this phenomenon baffled for long the community of the large-scale structure, it is now 
generally accepted that the cosmic web, anisotropic large-scale tidal environments surrounding DM halos \citep{bon-etal96}, 
must be mainly responsible for the deviation of the simple prediction of the excursion set theory on the halo growths from the reality 
\citep[e.g.,][]{san-etal07,hah-etal09,wan-etal11,zomg1,toj-etal17,yan-etal17,mus-etal18,MK19,ram-etal19}. 
Thus, a key to understanding the halo assembly bias is to figure out what aspect of the anisotropic tidal fields affects the halo growths. 

The cosmic web is further classified into four different types each of which has a distinct geometrical shape and dimension:  
zero dimensional knots, one dimensional filaments, two dimensional walls and three dimensional voids \citep{hah-etal07}. 
Among them, the most anisotropic web-type, the filament, turned out to embed the majority of DM halos \citep[e.g.,][]{gan-etal19} 
which were believed to grow via the preferential merging and accretion of satellites along the narrow one-dimensional 
channels \citep[e.g.,][]{wes-etal95,PB02,ver-etal11}.  
A recent numerical work of \citet{zomg1} based on a high-resolution N-body simulation, however, revealed that the motions 
of satellites confined in the filamentary environments could have opposite effects on the growths of galactic halos, depending on 
the filament thickness \citep[see also][]{GP16}. If multiple fine filaments cross one another at some nodes, the radial motions of the satellites 
along the filaments facilitate their infalls onto the galactic halos located at the nodes, enhancing the growths of the hosts.  
Whereas, in the bulky filaments thicker than the sizes of the constituent galactic halos, the satellites preferentially move in the tangential 
directions orthogonal to the filament axes, which lead to the deterrence of the satellite infalls and the retarded growths of their 
hosts. Quantifying the filament thickness in terms of the ellipticity of the surrounding large-scale structure and incorporating it into the conditions 
for the halo formation, \citet{zomg1} proposed a new extension of the excursion set theory which could accommodate the opposite effects of the 
large-scale tidal environments on the growths of the galactic halos \citep[see also][]{zomg3}. 

Motivated by the insightful work of \citet{zomg1}, several attempts were made to improve their model by incorporating more realistic 
conditions from the halo growths or by extending the model to the larger scales or to the other web types \citep{laz-etal17,mus-etal18,lee19}. 
For instance, \citet{lee19} introduced a new concept of the "tidal coherence" for a quantitative explicit description of the filament thickness, 
suggesting that bulky thick (multiple fine) filaments should be outcomes of the highly coherent ({\it incoherent}) tides defined as the 
strong (weak) alignments between the first eigenvectors corresponding to the larges eigenvalues of the tidal fields smoothed on two widely 
separated scales.  With the numerical analysis on the cluster scales, \citet{lee19}  indeed found that the radial (tangential) motions of the 
infall-zone satellites around host clusters are obstructed (facilitated) by the highly coherent tides, which implies that the halo growth sensitively 
depends on the degree of the tidal coherence. 

Yet, the prime focus of \citet{lee19} was the future evolution of the cluster halos rather than their past evolutions, dealing with the infall-zone 
satellites which have yet to fall into the halos. It is necessary to treat the real satellites for the investigation of the effect of the tidal coherence on the 
past growths of the DM halos. Besides, the original definition of the tidal coherence in terms only of the first eigenvector direction may 
neglect the possibilities that the coherence in the second and third eigenvector directions corresponding to the second largest and smallest 
eigenvalues are not evinced by the coherence in the first eigenvector direction and that the simultaneous coherence of the tides in multiple 
eigenvector directions may have different effects on the halo growths. 
 
In this Paper, we attempt to incorporate the multi-dimensional aspect of the tidal coherence into the idea of \citet{lee19} and to explore how 
it affects the halo growths by measuring a correlation between the mass-loss evolution of the halo satellites and the multi-dimensional 
tidal coherence. In Section \ref{sec:data} the definition of the multi-dimensional tidal coherence as well as the description of the 
numerical data sets utilized for this analysis are presented. In Sections \ref{sec:1d}-\ref{sec:3d}, the effects of the simultaneous coherence 
of the tides along one, two and three eigenvector directions on the subhalo mass-loss evolutions are presented. 
In Section \ref{sec:sum} the final results are summarized and its implication on the halo assembly bias is discussed. 
Throughout this analysis, we will assume a concordance cosmology with initial conditions prescribed by the Planck result \citep{planck13}. 

\section{Dependence of the Satellite Mass-Loss on the Tidal Coherence}

\subsection{Tidal Coherence as a Multi-Component Array}
\label{sec:data}

For this analysis, we utilize the catalog of the Rockstar halos \citep{rockstar} and density field at $z=0$ retrieved from the website of the Small 
MultiDark Planck simulation\footnote{https://www.cosmosim.org}\citep[SMDPL,][]{smdpl}, a DM-only N-body simulation performed on a periodic 
box of linear size $400\,h^{-1}$Mpc,  containing $3840^{3}$ DM particles of individual mass $m_{p}=9.63\times 10^{7}\,h^{-1}\,M_{\odot}$ for the 
Planck cosmology \citep{planck13}.  
The catalog contains both of the distinct halos and the subhalos, which can be distinguished by their parent ID (pId): The former 
has pId$=-1$ while the pId of the latter is nothing but the ID of its parent halo, a least massive distinct halo which gravitationally hosts the latter. 
Selecting as the hosts the massive cluster-size distinct halos in the mass range of $(1-3)10^{14}\,h^{-1}\,M_{\odot}$, 
we identify their subhalos whose pId's match their ID's. 

For each subhalo belonging to each host, we determine the ratio, $\xi_{m}\equiv M_{\rm vir}/M_{\rm acc}$, of its virial mass, $M_{\rm vir}$, 
to its accretion mass, $M_{\rm acc}$, defined as the subhalo mass at the moment of its accretion to its host.  The majority of the subhalos are  
to lose their masses after their infalls via various processes like the tidal stripping/heating and dynamical frictions \citep{bos-etal05}, 
for which cases we expect $\xi_{m}<1$. The lower value of $\xi_{m}$ below unity indicates that the given subhalo must have experienced 
the severe mass-loss processes for longer time after the infall. Yet, in some rare occasions, the subhalos can gain masses through merging 
inside the hosts for which case $\xi_{m}$ can exceed unity. From here on, two terms, {\it subhalos} and {\it satellites}, will be interchangeablly 
used to refer to the non-distinct Rockstar halos gravitationally bound to some larger distinct halos. 

As done in \citet{lee19}, we compute the tidal field, $T_{ij}({\bf x})$, from the density field defined on the $512^{3}$ grid points, 
$\rho({\bf x})$, by taking the following steps:
(i) Calculating the density contrast field as $\delta({\bf x})\equiv (\rho({\bf x})-\bar{\rho})/\bar{\rho}$ where $\bar{\rho}$ is the mean density 
averaged over the grid points. 
(ii) Performing the Fourier transformation of $\delta({\bf x})$ into $\tilde{\delta}({\bf k})$.
(iii) Smoothing the density field in the Fourier space with a Gaussian filter on the scale of $R_{f}=30\,h^{-1}$Mpc as 
$\tilde{\delta}_{s}({\bf k})\equiv \tilde{\delta}({\bf k})\exp(-k^{2}R^{2}_{f}/2)$.
(iv) Computing the Fourier amplitude of the tidal field as $\tilde{T}_{ij}\equiv k_{i}k_{j}\tilde{\delta}_{s}({\bf k})/k^{2}$.
(v) Performing the inverse Fourier transformation of $\tilde{T}_{ij}({\bf k})$ into $T_{ij}({\bf x})$. 
At the grid point, ${\bf x}_{h}$, where each of the selected hosts is located, we diagonalize $T_{ij}({\bf x}_{h})$ to find a set of three eigenvalues 
$\{\lambda_{i}\}_{i=1}^{3}$ (with a decreasing order) and the corresponding eigenvectors $\{{\bf e}_i\}_{i=1}^{3}$. Then, we repeat the whole 
process but with a smaller filtering scale of $R^{\prime}_{f}=2\,h^{-1}$Mpc to obtain a new set of $\{\lambda^{\prime}_{i}\}_{i=1}^{3}$ and 
$\{{\bf e}^{\prime}_i\}_{i=1}^{3}$.

As mentioned in Section \ref{sec:intro}, the tidal coherence, $q$, was originally defined as $q\equiv \vert{\bf e}_{1}\cdot{\bf e}^{\prime}_{1}\vert$
\citep{lee19}. In the current work, we redefine $q$ as a multi-component array as  
\begin{equation}
\label{eqn:multi_q}
q_{i}=\vert{\bf e}_{i}\cdot{\bf e}^{\prime}_{i}\vert\,  \qquad {\rm for}\ {\rm each}\ i\in \{1,2,3\}\, .
\end{equation}
If $q_{i}$ is equal to or higher than $0.9$ (lower than $0.2$) at a given region, the tides is said to be highly coherent ({\it incoherent}) along the 
$i$th eigenvector direction at the region. A critical question to which we would like to find an answer in the following Subsections is whether or not 
the subhalos located in the regions where the tides are highly coherent or {\it incoherent} in different eigenvector directions exhibit different 
mass-loss evolutions.

\subsection{One Dimensional Dependence}\label{sec:1d}

In this Subsection, we are going to study how the mean value of the subhalo virial-to-accretion mass ratios depends on each of the three 
components of the tidal coherence, $\{q_{i}\}_{i=1}^{3}$, calling it one-dimensional (1D) dependence of the subhalo mass-loss evolution on 
the tidal coherence. 
We first divide the sample of the selected host halos into two subsamples: One contains those hosts surrounded by the tides highly coherent along 
the first eigenvector direction, satisfying the condition of $q_{1}\ge 0.9$. The other consists of those surrounded by the tides not so strongly coherent 
along the first eigenvector direction with $q_{1}< 0.9$. Table \ref{tab:1d} lists the mean masses ($\langle M_{h}\rangle$) and numbers ($N_{h}$) 
of the hosts contained in each subsample. 
As can be seen, although the latter subsample (i.e., $q_{1}< 0.9$) contains three times larger number of hosts, no significant difference in 
$\langle M_{h}\rangle$ between the two subsamples is noted, which assures that if the values of $\langle\xi_{m}\rangle$ from the two subsamples 
are significantly different from each other, then it should not be ascribed to the mass difference.

For each host contained in each subsample, we select only those subhalos which experienced the {\it mass-loss} process, i.e., $\xi_{m}< 1$, 
excluding those few subhalos which experienced the mass-gain process, $\xi_{m}\ge 1$.  
Then, we calculate the mean virial-to-accretion mass ratio, $\langle \xi_{m}\rangle$, averaged over the selected subhalos of the hosts contained 
in each subsample. The errors, $\sigma_{\xi_{m}}$, in the measurement of $\langle\xi_{m}\rangle$, is calculated as its standard deviation as 
$\sigma_{\xi_{m}}\equiv [\langle(\xi_{m}-\langle\xi_{m}\rangle)^{2}\rangle/(N_{\rm sub}-1)]^{1/2}$ where $N_{\rm sub}$ is the total number of the 
subhalos of the hosts contained in each subsample.  

\begin{figure}
\begin{center}
\includegraphics[scale=0.9]{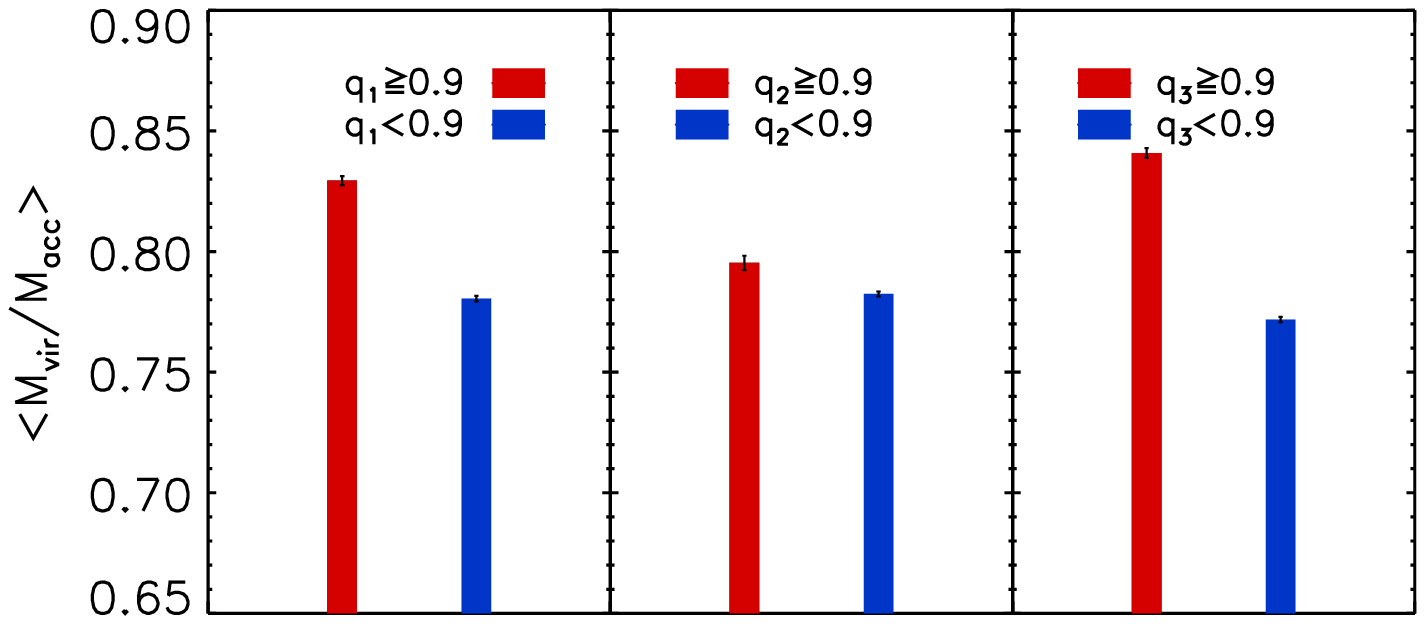}
\caption{Mean virial-to-accretion mass ratios of the subhalos belonging to the hosts surrounded by the 
tides highly coherent along one of three eigenvector directions as red bar (major, intermediate 
and third eigenvector directions in the left, middle and right panels, respectively.) 
In each panel, the complement case of the tides not so highly coherent along the same direction 
is plotted as blue bar.}
\label{fig:1d}
\end{center}
\end{figure}

Figure \ref{fig:1d} plots the values of $\langle\xi_{m}\rangle$ from the two subsamples with $q_{1}\ge 0.9$ and $q_{1}< 0.9$ as thick red and 
blue bars, respectively, with the associated errors $\sigma_{\xi_{m}}$  in its left panel, explicitly demonstrating that the former yields a significantly 
higher value of $\langle\xi_{m}\rangle$ than the latter. This trend implies that the satellites located in the regions surrounded by the tides highly 
coherent along the first eigenvector direction experience less severe mass-loss evolution after their infalls onto their hosts than the other 
counterparts with $q_{1}< 0.9$.  Based on the insights from \citet{lee19}, we put forth the following explanation to understand this phenomenon: 
As the satellites surrounded by highly coherent tides along the first eigenvector direction develop velocities in the tangential 
direction, which deter their infalls onto the hosts, reducing the amount of time during which the subhalos are exposed to the effects of the tidal 
stripping/heating or dynamical fraction inside their hosts. 

Repeating the above procedure but with the subsamples obtained by contraining the value of $q_{2}$ ($q_{3}$) instead of $q_{1}$ with 
the same threshold of $0.9$, we also investigate how $\langle\xi_{m}\rangle$ differs between the cases of $q_{2}\ge 0.9$ and 
$q_{2}<0.9$ ($q_{3}\ge 0.9$ and $q_{3}<0.9$). 
The middle (right) panel of Figure \ref{fig:1d} plots the same as the left panel but for the case that the subsample is divided by imposing the 
threshold condition on the value of $q_{2}$ ($q_{3}$).  As can be seen, the subhalos of the hosts located in the regions with $q_{2}\ge 0.9$ 
($q_{3}\ge 0.9$) yield a larger value of $\langle\xi_{m}\rangle$ than those with $q_{2}< 0.9$ ($q_{3}< 0.9$), the same trend as that shown in the 
left panel of Figure \ref{fig:1d}. Note, however, that the larger (smaller) difference in $\langle \xi_{m}\rangle$ between the two subsamples are 
found for the case that the threshold condition is imposed on the value of $q_{3}$ ($q_{2})$ rather than on the value of $q_{1}$. 

To see whether or not this difference in $\langle\xi_{m}\rangle$ witnessed in Figure \ref{fig:1d} is a secondary effect induced by any differences in  
the local density ($\delta$) or ellipticity ($e$) between the two subsamples, we determine the values $\delta$,and $e$ at the grid point of 
each host. The three tidal eigenvalues, $\{\lambda^{\prime}_{i}\}_{i=1}^{3}$ on the scale of $2\,h^{-1}$Mpc obtained in 
Subsection \ref{sec:data} is used to calculate $\delta$ and $e$: $\delta=\sum_{i=1}^{3}\lambda^{\prime}_{i}$, and 
$e\equiv [(1+\delta)^{-1}\sum_{i<j}(\lambda^{\prime}_{i}-\lambda^{\prime}_{j})^{2}]^{1/2}$ .  This definition of $e$, was 
devised by \citet{ram-etal19} to eliminate any correlation between $e$ and $\delta$. 

\begin{figure}
\begin{center}
\includegraphics[scale=0.9]{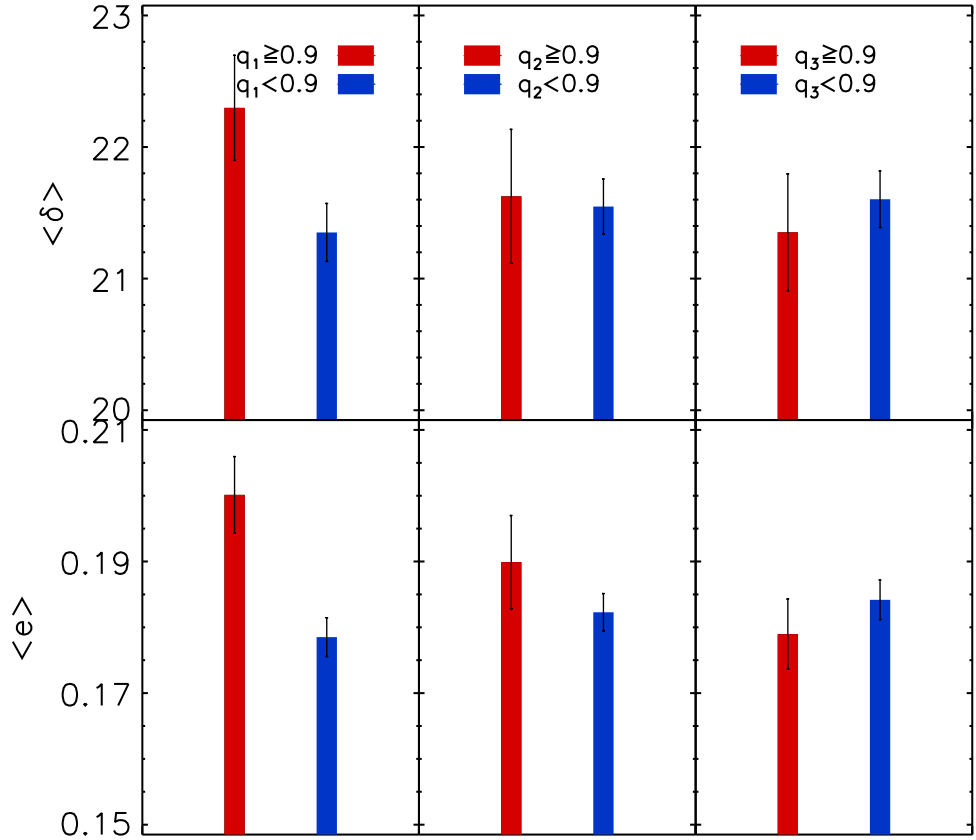}
\caption{Mean values of the density contrast and ellipticity averaged over the regions surrounded by the tides highly 
coherent along one of three eigenvector directions as red bar in the top and bottom panels, respectively.
In each panel, the blue bar correspond to the complement case. }
\label{fig:den_1d}
\end{center}
\end{figure}

Taking the mean values, $\langle\delta\rangle$ and $\langle e\rangle$, averaged over all hosts contained in each of the subsamples, we 
plot them in the top and bottom panels of Figure \ref{fig:den_1d}, respectively. 
As can be seen, when the value of $q_{2}$ or $q_{3}$ are constrained by using a threshold of $0.9$, no significant differences are found in 
$\langle\delta\rangle$ and $\langle e\rangle$ between the two subsamples. 
Whereas, the subsample with $q_{1}\ge 0.9$ is found to have substantially larger values of $\langle\delta\rangle$ and $\langle e\rangle$ 
than the other  subsample with $q_{1}<0.9$.  That is, the regions surrounded by the tides highly coherent along the first eigenvectors 
tend to be more overdense and more anisotropic due to the simultaneous compression of matter along the coherent first eigenvector direction.  
This result brings out a suspicion that the higher value of $\langle\xi_{m}\rangle$ found in the subsample with $q_{1}\ge 0.9$ may be 
caused by the higher values of $\langle\delta\rangle$ and $\langle e\rangle$. 

\begin{figure}
\begin{center}
\includegraphics[scale=0.9]{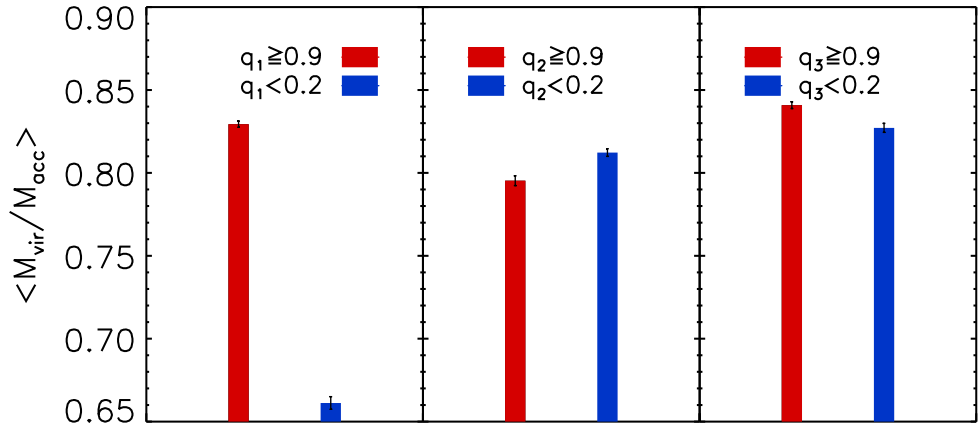}
\caption{Mean virial-to-accretion mass ratios of the subhalos belonging to the hosts surrounded by the 
tides highly coherent ({\it incoherent}) along one of three eigenvector directions as red (blue) bar.}
\label{fig:1d_hl}
\end{center}
\end{figure}

\begin{figure}
\begin{center}
\includegraphics[scale=0.9]{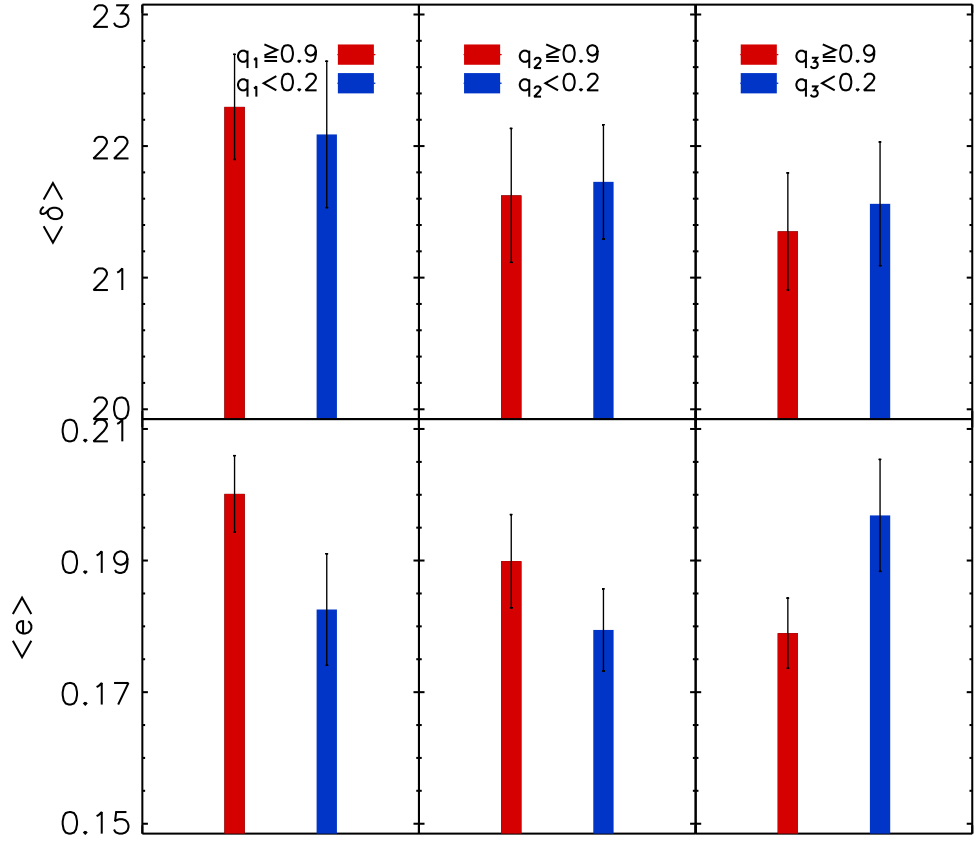}
\caption{Mean values of the density contrast and ellipticity averaged over the regions surrounded by the tides highly 
coherent ({\it incoherent}) along one of three eigenvector directions as red (blue) bar
in the top and left panels, respectively.}
\label{fig:den_1d_hl}
\end{center}
\end{figure}

Now that the tides highly coherent along the eigenvector direction are found to have an obstruction effect on the satellite infalls, 
the next quest is to investigate whether the tides highly {\it incoherent} along any eigenvector direction have the opposite effect or not.  
For this quest,  we use two thresholds: an upper-bound threshold of $0.2$ and a lower-bound threshold of $0.9$ to construct two subsamples 
(i.e., $q_{i}\ge 0.9$ and $q_{i}< 0.2$ for each $i\in \{1,2,3\}$) and then conduct the same analysis. 
Figures \ref{fig:1d_hl}-\ref{fig:den_1d_hl} plot the same as Figures \ref{fig:1d}-\ref{fig:den_1d}, respectively, but with the conditions of 
$q_{i}\ge 0.9$ and $q_{i}< 0.2$ instead of $q_{i}\ge 0.9$ and $q_{i}<0.9$. 
The left panel of Figure \ref{fig:1d_hl} reveals that the difference in $\langle\xi_{m}\rangle$ between the two subsamples obtained by 
putting two thresholds of $0.9$ and $0.2$ on the value of $q_{1}$ is larger than that by putting one threshold of $0.9$. 
This result indicates that the tides highly {\it incoherent} along the first eigenvector direction indeed have the opposite effect on the satellite 
infalls: it facilitates the satellite infalls onto the hosts, leading them to undergo the more severe mass-loss evolution after the infalls. 
Meanwhile, the left panel of Figure \ref{fig:den_1d_hl} shows that the difference in $\langle\delta\rangle$ and $\langle e\rangle$ between the two 
subsamples obtained by putting two thresholds of $0.9$ and $0.2$ on the value of $q_{1}$ is smaller than that by using one threshold of $0.9$, 
which proves that the larger values of $\langle\delta\rangle$ and $\langle e\rangle$ are not mainly responsible for the more severe 
mass-loss evolution of the subhalos found from the subsample with $q_{1}\ge 0.9$.  

It is interesting, however, to discover in the right panel of Figure \ref{fig:1d_hl} that the tides highly {\it incoherent} along the third eigenvector 
direction does not have the expected opposite effect, compared to that coherent along the same direction. 
The difference in $\langle\xi_{m}\rangle$ between the subsamples obtained by putting two thresholds of $0.9$ and $0.2$ on $q_{3}$ is smaller 
than that between the subsamples obtained by putting one threshold of $0.9$ on $q_{3}$. This result indicates that the tides highly {\it incoherent} 
along the third eigenvector direction have an obstructing effect on the satellite infalls rather than facilitating it unlike the tides highly {\it incoherent} 
along the first eigenvector direction. This phenomenon may be closely linked with the larger mean ellipticity, $\langle e\rangle$, found in the 
subsample with $q_{3}<0.2$ than in the subsample with $q_{3}<0.9$, shown in the right panel of Figure \ref{fig:den_1d_hl}. 
The tides highly {\it incoherent} along the third eigenvector direction can increase the tidal anisotropy of a region, which in turn 
makes it harder for the satellites in the region to fall onto their hosts. As the satellite infalls are deterred, they must go through less severe mass-loss 
evolution after the infalls till the present epochs.  Note also in the middle panels of Figures \ref{fig:1d}-\ref{fig:den_1d_hl} that the tidal coherence 
measured along the second eigenvector direction have the weakest effect on the subhalo mass-loss evolution, showing no significant differences 
in $\langle\xi_{m}\rangle$, $\langle\delta\rangle$, and $\langle e\rangle$ among three samples with $q_{2}\ge 0.9$, $q_{2}< 0.9$ and $q_{2}<0.2$. 

\subsection{Two Dimensional Dependence}\label{sec:2d}

Now that the surrounding tides coherent along different eigenvector directions are found to have different effects on the mass-loss evolution of the 
subhalos, we would like to explore the effects of the tides coherent simultaneously along two eigenvector directions. Since the tidal coherence 
measured along the second eigenvector direction is found to have the weakest effect on the subhalo mass-loss evolution in Subsection 
\ref{sec:1d}, we will focus on the tidal coherence measured simultaneously along the first and third eigenvector directions (i.e., $q_{1}$ and $q_{3}$) 
in this Subsection. 

\begin{figure}
\begin{center}
\includegraphics[scale=0.9]{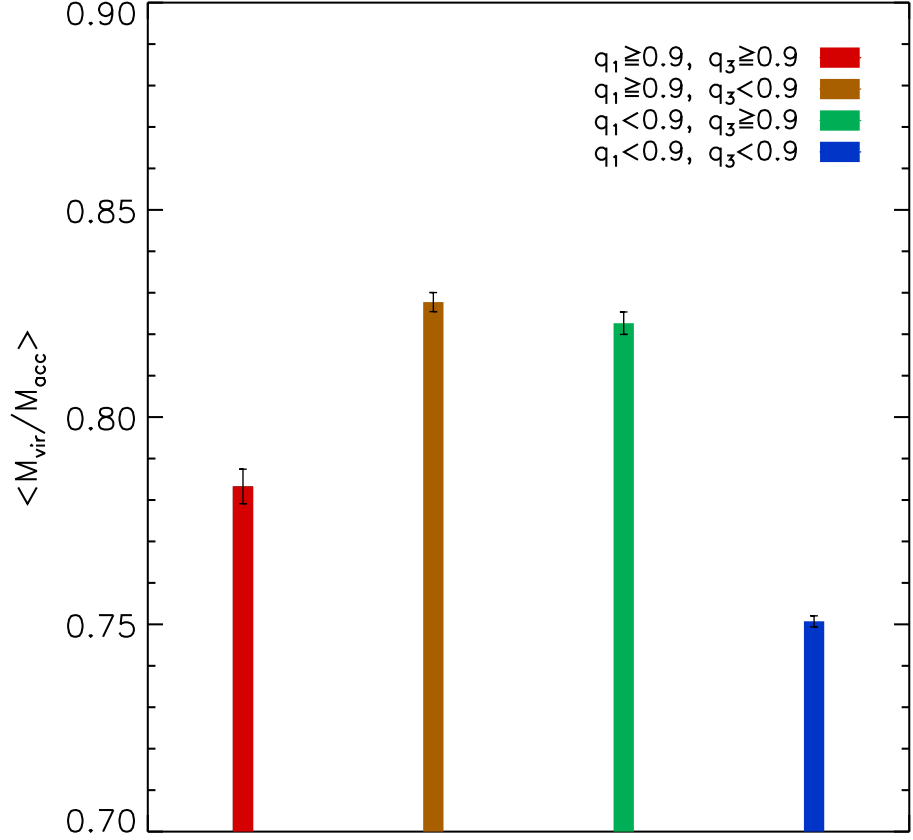}
\caption{Mean virial-to-accretion mass ratios of the subhalos belonging to the hosts surrounded by the 
tides highly coherent along both of the first and third (thick red bar), along the first but not along the third 
(thick ocher bar), highly coherent along the third but not along the first (thick green bar), and highly 
coherent along none of the first and third eigenvector directions (thick blue bar).}
\label{fig:2d}
\end{center}
\end{figure}

\begin{figure}
\begin{center}
\includegraphics[scale=0.9]{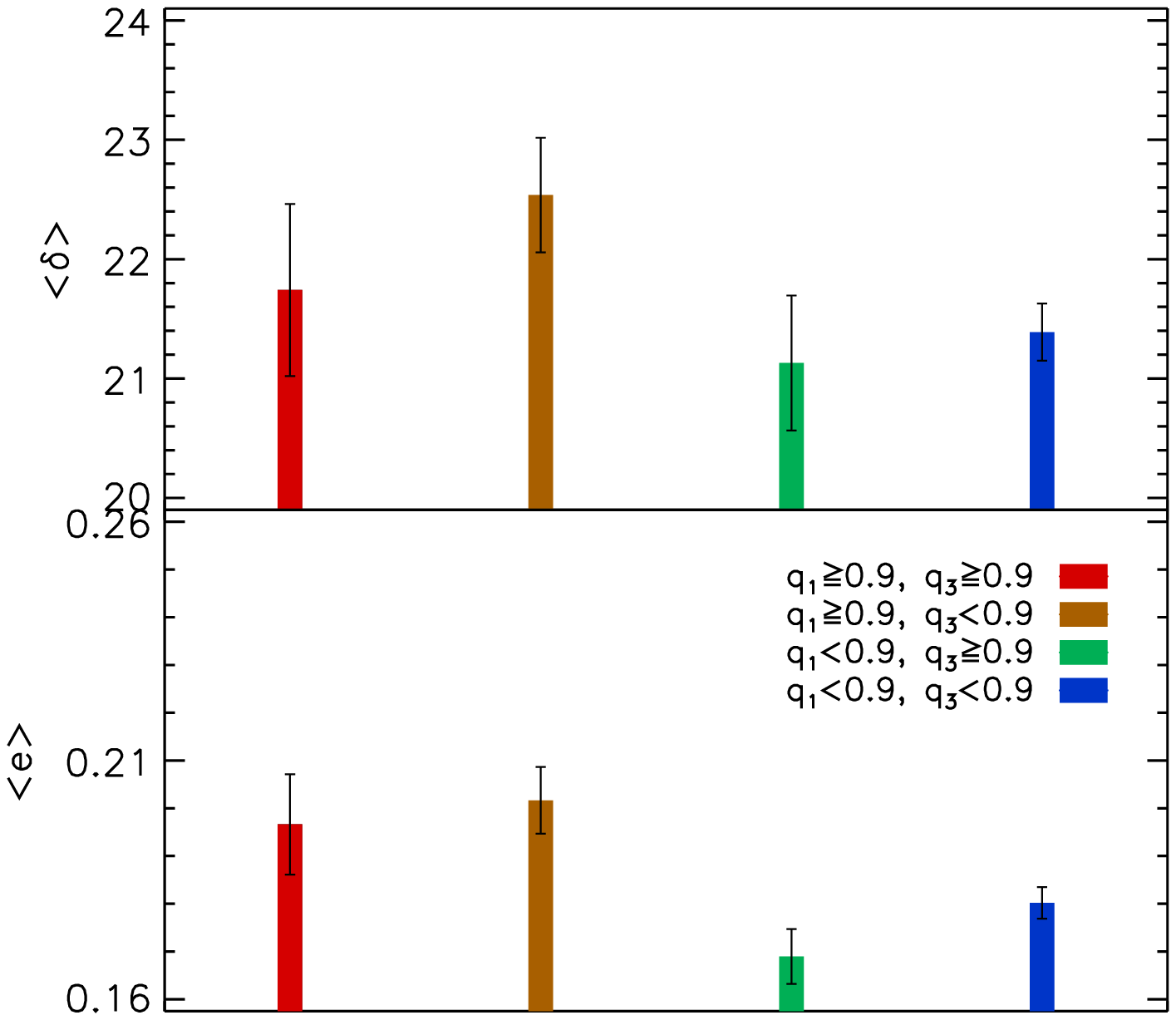}
\caption{Mean values of the density contrast and ellipticity averaged over the regions surrounded by the tides 
for the four different cases described in the caption of Figure \ref{fig:2d} in the top and bottom panels, respectively.}
\label{fig:den_2d}
\end{center}
\end{figure}

We first separate the selected host halos into four subsamples by simultaneously constraining the values of $q_{1}$ and $q_{3}$ with a single 
threshold of $0.9$ (see Table \ref{tab:2d}). Then, we calculate ($\langle \xi_{m}\rangle,\ \sigma_{\xi_{m}}$), ($\langle\delta\rangle,\ \sigma_{\delta}$) 
and ($\langle e\rangle,\ \sigma_{e}$) by taking the same steps described in Subsection \ref{sec:1d} for each of the four subsamples, the results of 
which are displayed in Figures \ref{fig:2d}-\ref{fig:den_2d}.  
As can be seen in Figure \ref{fig:2d}, the two subsamples satisfying the conditions of $(q_{1}\ge 0.9,\ q_{3}< 0.9)$ and $(q_{1}<0.9,\ q_{3}\ge 0.9)$ 
yield significantly higher values of $\langle\xi_{m}\rangle$ than the other two subsamples. A crucial implication of this result is that the tides highly 
coherent along the first (third) eigenvector direction but not along the third (first) eigenvector directions have a stronger obstructing effect on 
the satellite infalls than the tides highly coherent along both of the first and third eigenvector directions.

It is interesting to see that while the two subsamples with of $(q_{1}\ge 0.9,\ q_{3}< 0.9)$ and $(q_{1}<0.9,\ q_{3}\ge 0.9)$ show no significant 
difference in the values of $\langle\xi_{m}\rangle$ and $\langle\delta\rangle$ from each other, a substantial difference in the value of 
$\langle e\rangle$ is found between them (see Figure \ref{fig:den_2d}): the regions surrounded by the tides highly coherent along the first 
eigenvector direction but not along the third ones are more anisotropic  than those surrounded by the tides highly 
coherent along the third eigenvector direction but not along the first ones. 
Given that the tidal anisotropy can also have an effect of obstructing the satellite infalls,  the larger value of $\langle\xi_{m}\rangle$ found 
from the subsample with $q_{1}\ge 0.9$ and $q_{3}< 0.9$ may be partly caused by its larger value of $\langle e\rangle$ than that from 
the subsample with $q_{1}< 0.9$ and $q_{3}\ge 0.9$.   
The lowest value of $\langle\xi_{m}\rangle$ is found from the subsample with $q_{1}< 0.9$ and $q_{2}< 0.9$, 
which indicates that the tides highly coherent along none of the first nor third eigenvector directions have the weakest obstructing 
and/or strongest facilitating effects of the satellite infalls. 

\begin{figure}
\begin{center}
\includegraphics[scale=1.0]{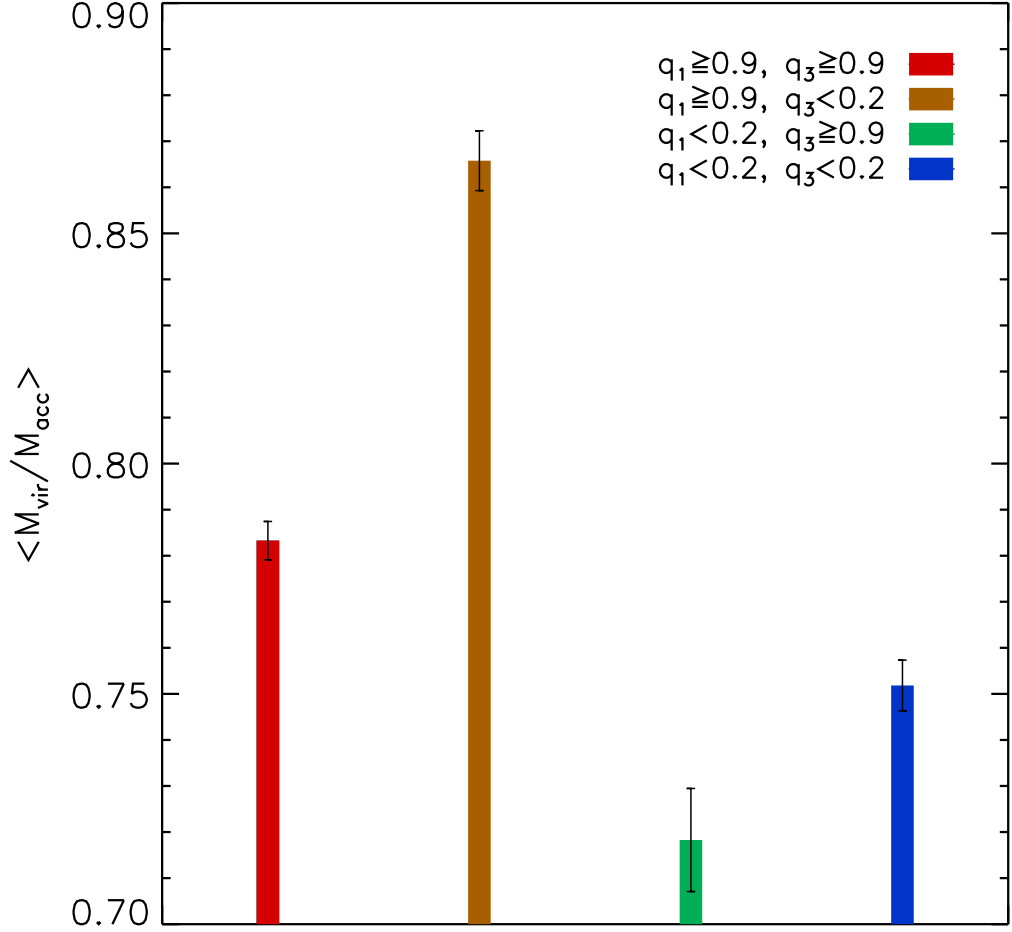}
\caption{Mean virial-to-accretion mass ratios of the subhalos belonging to the hosts surrounded by the 
tides highly coherent along both of the first and third (red bar), highly coherent along the first but 
highly {\it incoherent} along the third (ocher bar), highly coherent along the third but highly {\it incoherent} 
along the first (green bar), and highly {\it incoherent} along the first and third eigenvector directions 
(blue bar).}
\label{fig:2d_hl}
\end{center}
\end{figure}

\begin{figure}
\begin{center}
\includegraphics[scale=0.9]{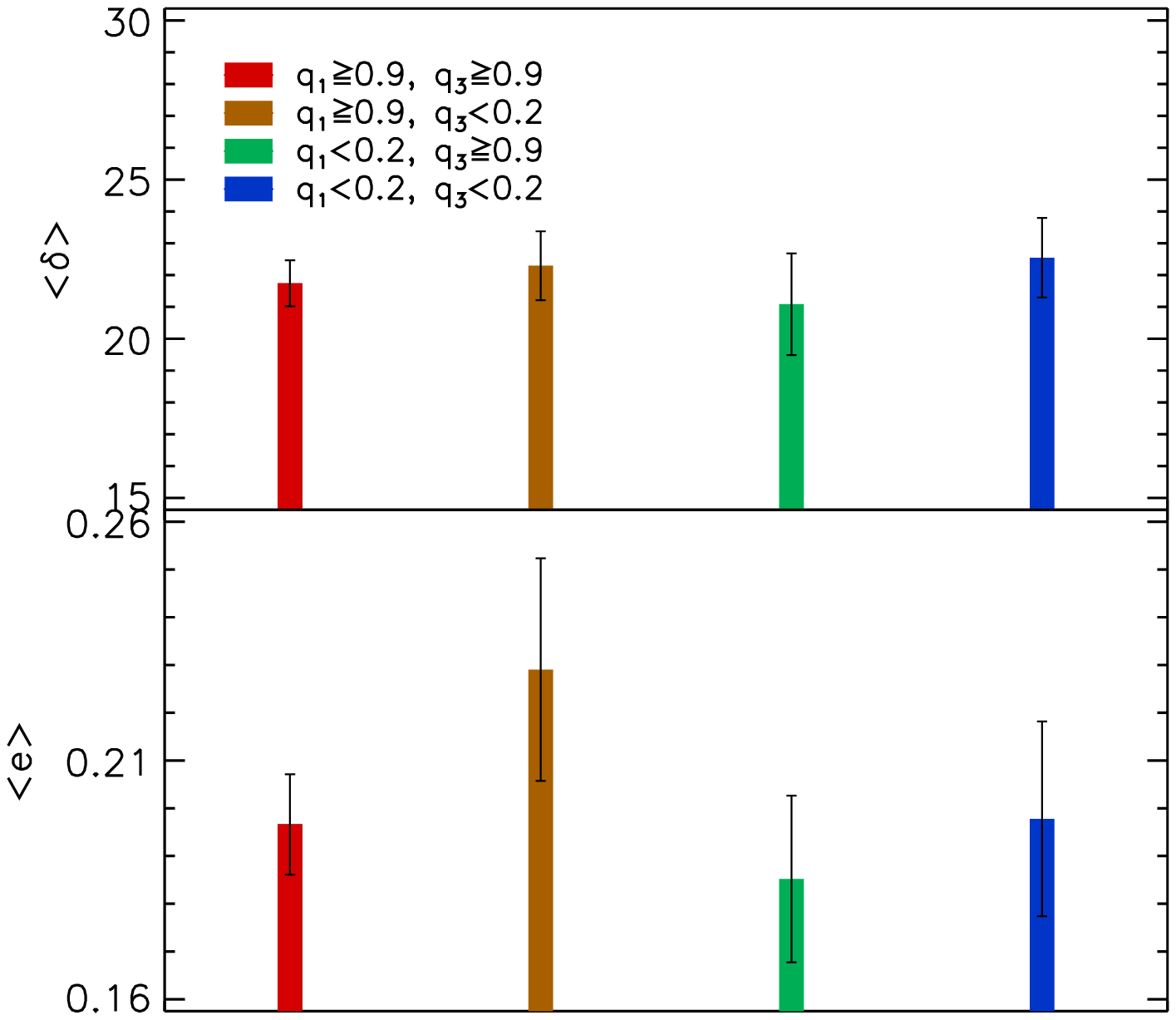}
\caption{Mean values of the density contrast and ellipticity averaged over the regions surrounded by the tides 
for the four different cases described in the caption of Figure \ref{fig:2d_hl} in the top and bottom panels, respectively.}
\label{fig:den_2d_hl}
\end{center}
\end{figure}

We also investigate the effect of the highly {\it incoherent} tides on the subhalo mass-loss evolution and on the local density and ellipticity 
as well by constraining the value of $q_{1}$ and $q_{3}$ with double thresholds of $0.9$ and $0.2$, the results of which are shown in 
Figures \ref{fig:2d_hl}-\ref{fig:den_2d_hl}. 
As can be seen in Figure \ref{fig:2d_hl}, the subsample with $q_{1}\ge 0.9$ and $q_{3}<0.2$ yields the highest value of 
$\langle\xi_{m}\rangle$ among the four, while its lowest value is found in the subsample with $q_{1}<0.2$ and $q_{3}\ge 0.9$. 
This result indicates that the tides highly coherent along the first eigenvector direction and highly {\it incoherent} along the third 
eigenvector direction are most effective in obstructing the satellite infalls, while the tides highly coherent along the third eigenvector 
direction and {\it incoherent} along the first eigenvector direction are most effective in facilitating the infalls among the four.  
Given that the subsample with ($q_{1}\ge 0.9,\ q_{3}<0.2$) yields the highest value of $\langle e\rangle$ among the four, 
the largest value of $\langle\xi_{m}\rangle$ from the subsample with $q_{1}\ge 0.9$ and $q_{3}<0.2$ should be partially caused 
by a larger value of $\langle e\rangle$. 

It is worth recalling that in Subsection \ref{sec:1d} the tides highly coherent only along the third eigenvector direction have been already found to 
obstruct the satellite infalls rather than facilitate them (see the right panel of Figure \ref{fig:1d}).  Nevertheless, if the tides are simultaneously 
incoherent along the first eigenvector direction, then the facilitating effect of the tidal {\it incoherence} along the first eigenvector direction 
seem to overwhelm the obstructing effect of the tidal coherence along the third eigenvector direction, according to the result shown in 
Figure \ref{fig:1d_hl}. In other words, it is the tidal {\it incoherence} along the first eigenvector direction that plays the most decisive dominant 
role of facilitating the satellite infalls, driving the largest amount of mass-loss of the subhalos in the post-infall stages. 

Meanwhile, the high coherence of the tides along the first eigenvector direction seems to be synergetic with its simultaneous {\it incoherence} 
along the third eigenvector direction (see Figure \ref{fig:2d_hl}). The subsample with ($q_{1}\ge 0.9,\ q_{2}<0.2$) yields the lowest value of 
$\langle\xi_{m}\rangle$ not only among the subsamples obtained by simultaneously constraining both of $q_{1}$ and $q_{3}$ but also 
among the subsamples obtained by constraining only one of three components of $\{q_{i}\}$ (see Figure \ref{fig:1d_hl}). Our interpretation is 
that the high tidal anisotropy associated with the tides highly {\it incoherent} along the third eigenvector direction tends to magnify the 
obstructing effect of the high tidal coherence along the first eigenvector direction. 

\subsection{Three Dimensional Dependence}\label{sec:3d}

Now that the simultaneous constraints of $q_{1}$ and $q_{3}$ uncovers the complex two-dimensional dependence of the subhalo mass-loss 
evolution on the tidal coherence, it should be legitimate to investigate how $\langle\xi_{m}\rangle$ depends on all of the three components 
of $\{q_{i}\}_{i=1}^{3}$, calling it three dimensional (3D) dependence of the subhalo mass-loss evolution on the tidal coherence. 
We first separate the host halos into eight subsamples by constraining simultaneously the values of $(q_{1},\ q_{2},\ q_{3})$ with a single 
threshold of $0.9$ (see Table \ref{tab:3d}). 

\begin{figure}
\begin{center}
\includegraphics[scale=1.0]{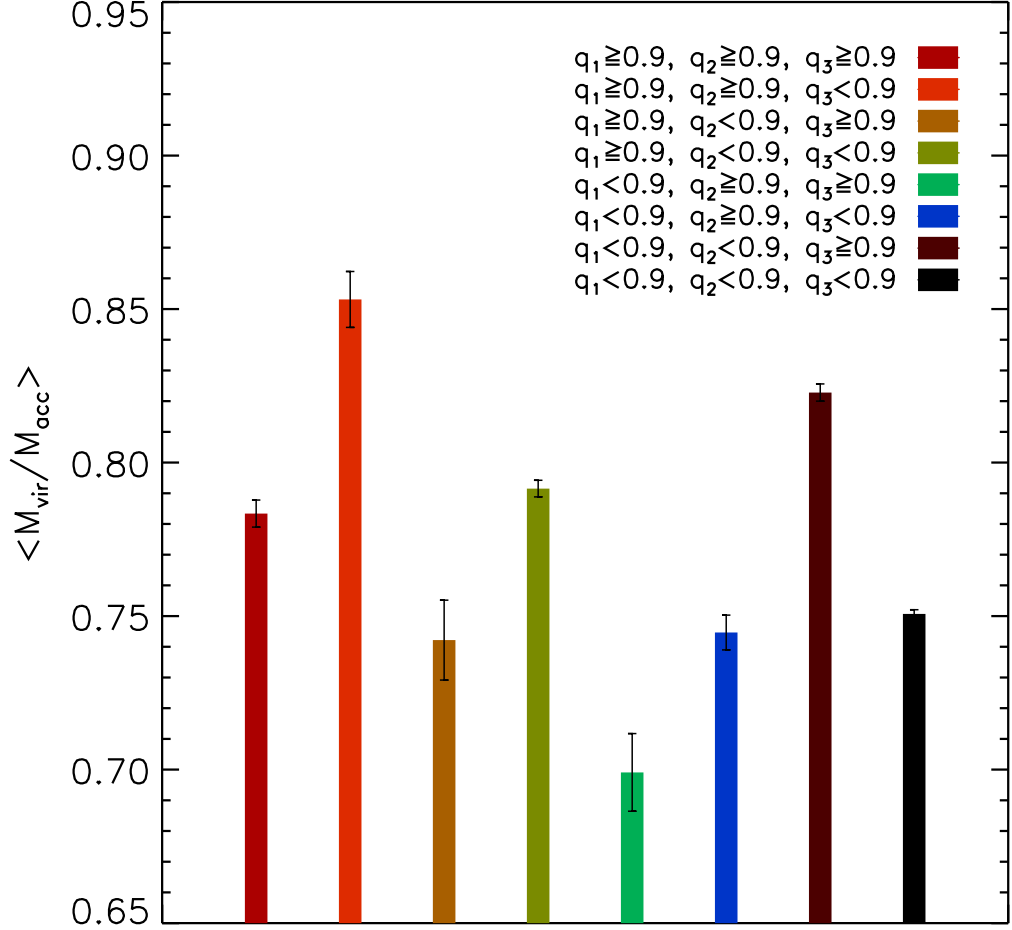}
\caption{Mean virial-to-accretion mass ratios of the subhalos belonging to the hosts surrounded by the tides highly 
coherent along all of the three eigenvector directions are plotted as carmine bar. The seven complement 
cases corresponding to the tides coherent along not all of the three eigenvector directions are plotted as different 
color bars.} 
\label{fig:3d}
\end{center}
\end{figure}

\begin{figure}
\begin{center}
\includegraphics[scale=0.9]{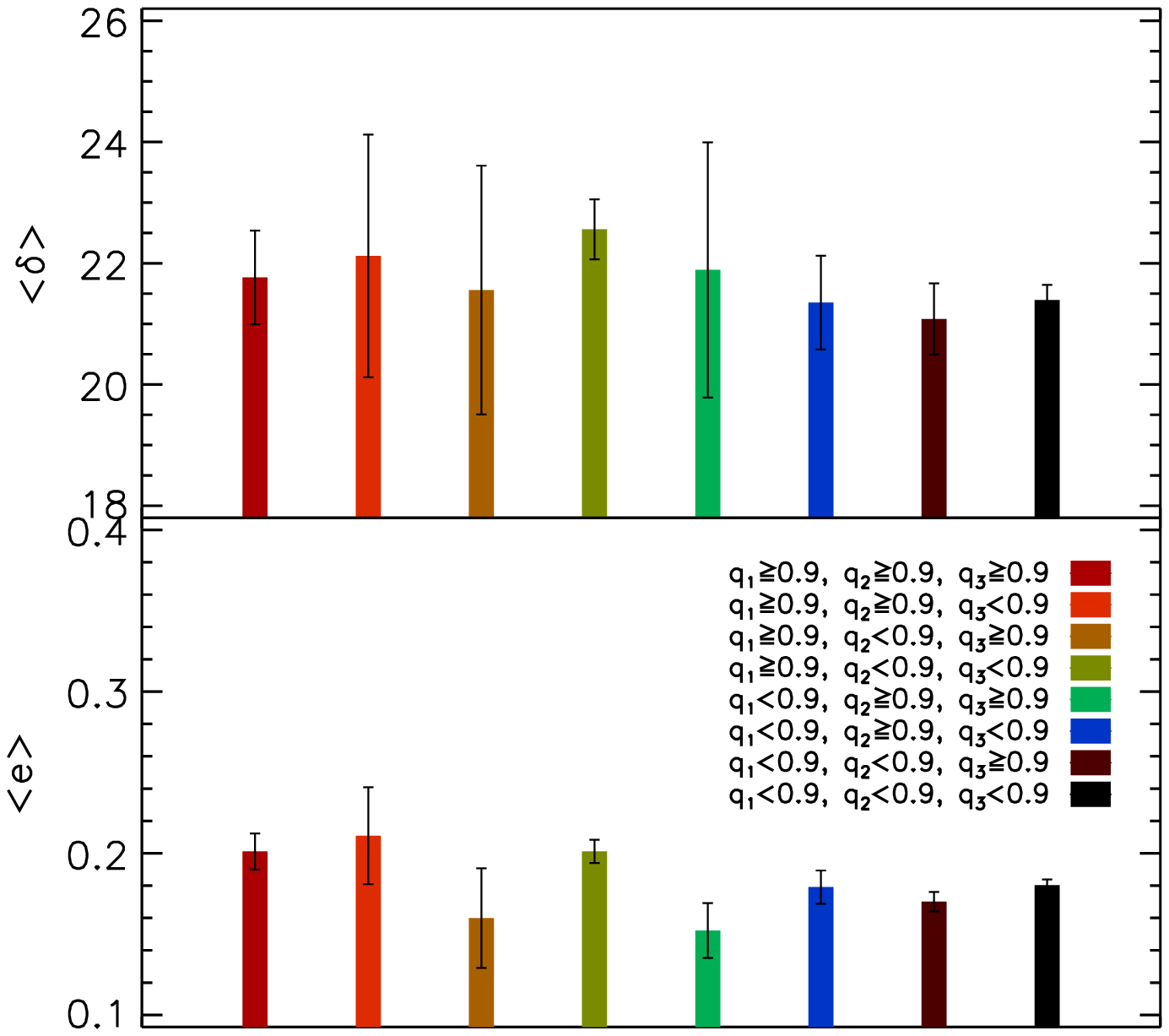}
\caption{Mean values of the density contrast and ellipticity averaged over the regions surrounded by the tides 
for the eight different cases described in the caption of Figure \ref{fig:3d} in the top and bottom panels, respectively.}
\label{fig:den_3d}
\end{center}
\end{figure}

Through the same procedure described in Subsection \ref{sec:1d}, we determine the values of ($\langle \xi_{m}\rangle,\ \sigma_{\xi_{m}}$), 
($\langle\delta\rangle,\ \sigma_{\delta}$) and ($\langle e\rangle,\ \sigma_{e}$), for each of the eight subsamples, which are plotted in 
Figures \ref{fig:3d}-\ref{fig:den_3d}.   
As can be seen in Figure \ref{fig:3d}, we find the highest and lowest values of $\langle\xi_{m}\rangle$ from the subsamples with 
$(q_{1}\ge 0.9, q_{2}\ge 0.9, q_{3}<0.9)$ and $(q_{1}< 0.9, q_{2}\ge 0.9, q_{3}\ge 0.9)$, respectively, among the eight. 
In Figure \ref{fig:den_3d} where the eight subsamples exhibit little difference in $\langle\delta \rangle$ but substantial difference 
in $\langle e\rangle$, we find the highest and lowest values of $\langle e\rangle$ from the same two subsamples, which implies that the large 
difference in $\langle e\rangle$ among the two subsamples should be linked with the large difference in $\langle\xi_{m}\rangle$. 

For the case that $q_{1}\ge 0.9$ and $q_{3}<0.9$, the simultaneous constraint of $q_{2}\ge 0.9$ gives the highest value of $\langle\xi_{m}\rangle$ 
(scarlet bar). Whereas, for the case that $q_{1}<0.9$ and $q_{3}\ge 0.9$,  the same constraint of $q_{2}\ge 0.9$ yields the opposite signal, i.e., 
the lowest value of $\langle\xi_{m}\rangle$ (green bar).  Note also that the subsample with $(q_{1}<0.9,\ q_{2}\ge 0.9,\ q_{3}<0.9)$ 
corresponding to the tides highly coherent only along the second eigenvector direction but not along the first and third ones yields relatively low 
value of $\langle\xi_{m}\rangle$. 
This result indicates that the effect of the high tidal coherence along the second eigenvector direction shifts from the obstruction to the facilitation 
of the satellite infalls,  depending on which eigenvector direction between the first and third the tides are simultaneously coherent. If the tides are 
highly coherent along none of the first and third eigenvector direction, then the high tidal coherence along the second eigenvector direction does not 
have a strong effect on the satellite infalls. 

It is interesting to see that the tides highly coherent along all of the three eigenvector directions (red bar) are less effective in obstructing the 
satellite infalls than the tides highly coherent along the first and second eigenvector directions but not highly coherent along the third eigenvector 
direction (scarlet bar). It is even not so effective in obstructing the satellite infalls as the tides highly coherent only along the third eigenvector 
direction but not along the first and second eigenvector direction (thick violet bar). Note also that the second highest value of 
$\langle\xi_{m}\rangle$ is found from the subsample with $q_{1}<0.9,\ q_{2}<0.9,\ q_{3}\ge 0.9$ (violet bar). 
Given that the mean ellipticity from this subsample is relatively low compared with the other seven cases (see Figure \ref{fig:den_3d}), this result 
implies that the net obstructing effect of the tides highly coherent only along the third but not along the first and second eigenvector directions may 
be stronger than that of the tides highly coherent only along the first and second eigenvector directions but not along the third eigenvector direction 
(scarlet bar). 

\begin{figure}
\begin{center}
\includegraphics[scale=0.9]{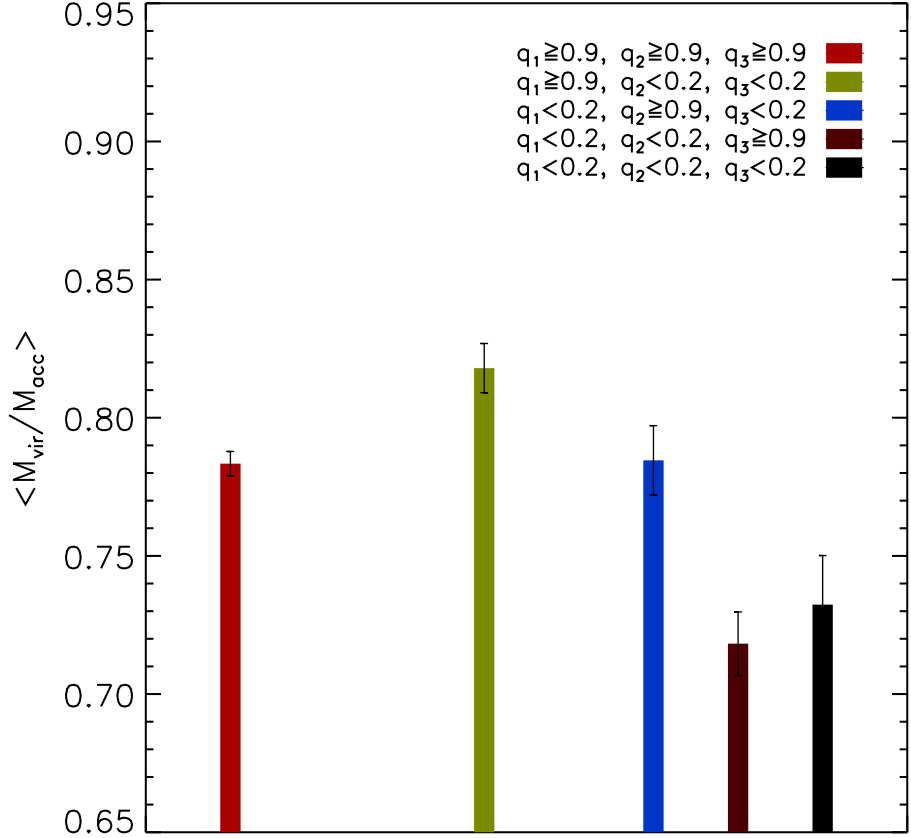}
\caption{Mean virial-to-accretion mass ratios of the subhalos belonging to the hosts surrounded by the tides highly 
{\it incoherent} along all of the three eigenvector directions are plotted as black bar. For comparison, the case of 
the tides coherent along all of the three directions are also plotted as carmine bar. The results from the three cases 
of the tides highly {\it incoherent} along two of the three eigenvector directions but highly coherent along the other 
directions are plotted as different color bars.}
\label{fig:3d_hl}
\end{center}
\end{figure}

\begin{figure}
\begin{center}
\includegraphics[scale=1.0]{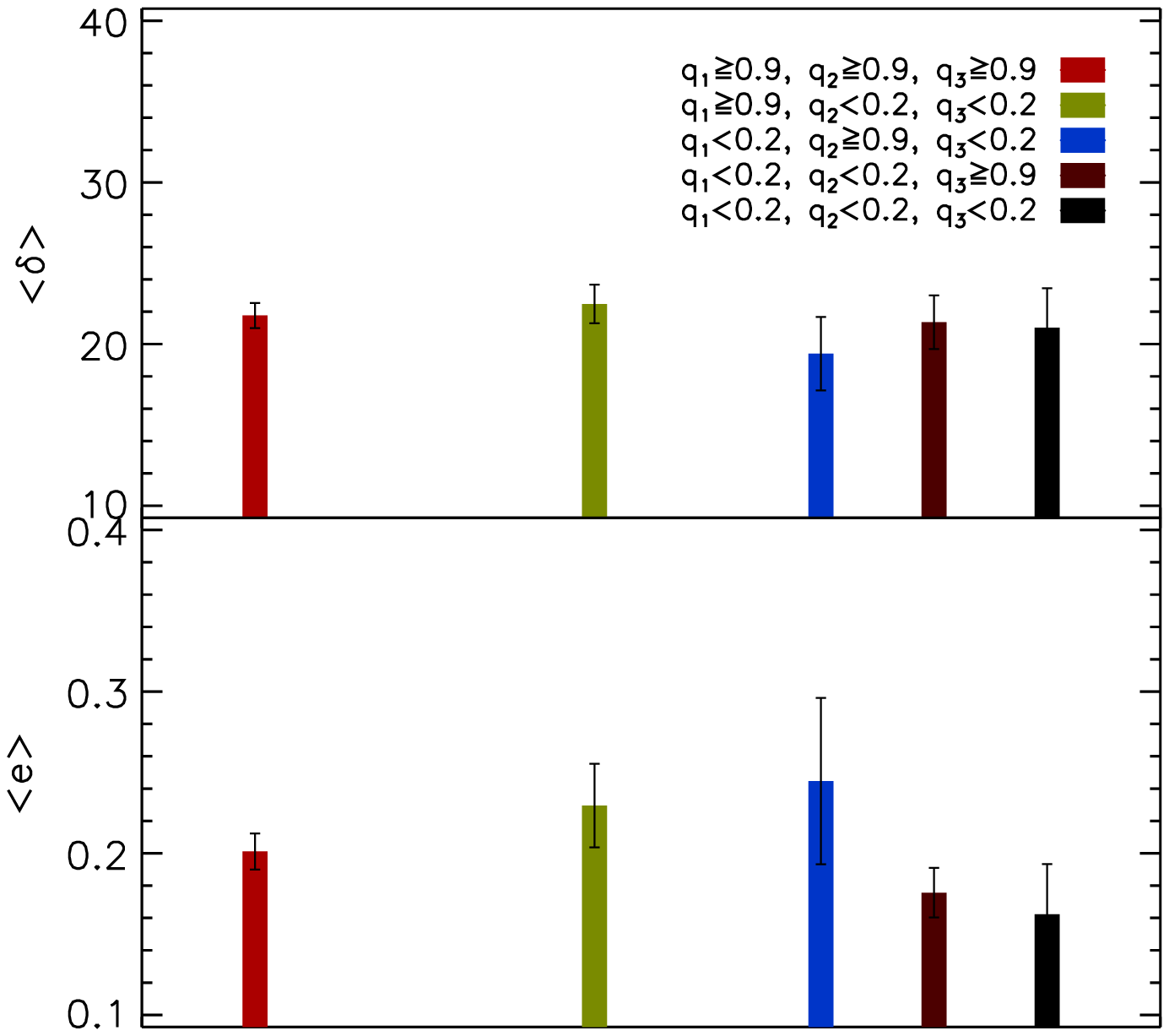}
\caption{Mean values of the density contrast and ellipticity averaged over the regions surrounded by the tides 
for the five different cases described in the caption of Figure \ref{fig:3d_hl} in the top and bottom panels, respectively.}
\label{fig:den_3d_hl}
\end{center}
\end{figure}

As done in Subsections \ref{sec:1d} and \ref{sec:2d}, we also investigate how the degree of the tidal {\it incoherence} measured along 
all of three eigenvector directions is linked with the subhalo mass-loss evolution, creating seven new subsamples by constraining simultaneously 
all of the three components, $(q_{1},\ q_{2},\ q_{3})$  with double thresholds of $0.9$ and $0.2$ (see Table \ref{tab:3d}): It turns out that no hosts 
satisfy the conditions of $(q_{1}\ge0.9,\ q_{2}< 0.2,\ q_{3}< 0.2)$, $(q_{1}<0.2,\ q_{2}\ge 0.9,\ q_{3}\ge 0.9)$ and 
$(q_{1}< 0.2,\ q_{2}\ge 0.9,\ q_{3}< 0.2)$, 
leaving three subsamples empty.
The values of $(\langle\xi_{m}\rangle,\ \sigma_{\xi_{m}})$, $(\langle\delta\rangle,\ \sigma_{\delta})$ and $(\langle e\rangle,\ \sigma_{e})$ 
obtained from the rest four non-empty subsamples as well as from the subsample with $(q_{1}\ge 0.9,\ q_{2}\ge 0.9,\ q_{3}\ge 0.9)$ are 
shown in Figures \ref{fig:3d_hl}-\ref{fig:den_3d_hl}.  

The subsample with $(q_{1}\ge 0.9,\ q_{2}< 0.2,\ q_{3}< 0.2)$ yields the highest value of $\langle\xi_{m}\rangle$ (olive green bar), while the lowest 
value (violet bar) is found from the subsample with $(q_{1}< 0.2,\ q_{2}< 0.2,\ q_{3}\ge 0.9)$.  Since the difference in $\langle e\rangle$ 
between the two subsamples is not so large enough to explain their difference in $\langle\xi_{m}\rangle$ (see Figure \ref{fig:den_3d_hl}), the 
different mean ellipticities between the two subsamples should not be the main cause of the significant difference in the mean virial-to-accretion 
mass ratios between them.  
The tides highly coherent along the first eigenvector direction but highly incoherent along the second and third eigenvector directions are much 
more effective in obstructing the satellite infalls than the tides highly coherent along the third eigenvector direction but highly incoherent along the 
first and second eigenvector directions. 

The comparison of the result shown in Figures \ref{fig:3d} and \ref{fig:3d_hl} reveals that the subsample with 
$(q_{1}<0.9,\ q_{2}\ge 0.9,\ q_{3}\ge 0.9)$ yield a lower value of $\langle\xi_{m}\rangle$ than the subsample with 
$(q_{1}<0.2,\ q_{2}< 0.2,\ q_{3}\ge 0.9)$. The tides highly coherent along the third eigenvector direction but highly {\it incoherent} along the first and 
second eigenvector direction are less effective in facilitating the satellite infalls than the tides highly coherent along the second and third eigenvector 
direction but not so highly coherent along the first eigenvector direction.

Another interesting fact revealed by the comparison between the two Figures is that the value $\langle\xi_{m}\rangle$ from the subsample with 
$(q_{1}<0.2,\ q_{2}\ge 0.9,\ q_{3}< 0.2)$ is as high as that from the subsample with $(q_{1}\ge 0.9,\ q_{2}\ge 0.9,\ q_{3}\ge 0.9)$. 
This result indicates that the tides highly coherent only along the second eigenvector direction but highly {\it incoherent} along the first and third 
eigenvector directions are as effective in obstructing the satellite infalls as the tides highly {\it coherent} along all of the three eigenvector directions. 
It is a rather surprising unexpected result since we have already found in Subsection \ref{sec:1d} that the tides highly coherent along the first 
eigenvector direction have an obstructing effect on the satellite infalls and that the tides highly {\it incoherent} along the same direction have the 
opposite effect, i.e., facilitating the satellite infalls.   
The slightly larger value of $\langle e\rangle$ from the subsample with $(q_{1}<0.2,\ q_{2}\ge 0.9,\ q_{3}< 0.2)$ 
than that from the subsample with $(q_{1}\ge 0.9,\ q_{2}\ge 0.9,\ q_{3}\ge 0.9)$ should be related to this puzzling phenomenon (see 
Figures \ref{fig:den_3d}-\ref{fig:den_3d_hl}). 
The tidal {\it incoherence} along the third eigenvector direction tends to increases the tidal anisotropy (i.e., mean ellipticity) which plays a role 
in increasing the value of $\langle\xi_{m}\rangle$, as shown in Subsection \ref{sec:1d}. The obstructing effect of the high tidal anisotropy caused 
by the tidal {\it incoherence} along the third eigenvector direction compensates the facilitating effect of the high {\it incoherence} of the tides 
along the first eigenvector direction. 

\section{Summary and Discussion}\label{sec:sum}

We have systematically studied the dependence of the subhalo mass-loss evolution on the multi-dimensional aspect of the tidal coherence by using 
the numerical datasets retrieved from the SMPDL \citep{smdpl}. For this study, we have quantified the subhalo mass-loss evolution in terms of 
the mean virial-to-accretion mass ratios averaged over the subahlos, and expressed the tidal coherence as an array of three numbers, 
$\{q_{i}\}_{i=1}^{3}$, 
where $q_{i}$ represents the alignments between the $i$th eigenvectors of the tidal fields smoothed on two widely separated scales of 
$2\,h^{-1}$Mpc and $30\,h^{-1}$Mpc. To eliminate the well known strong dependence of the subhalo mass-loss evolution on the masses of their 
hosts \citep{bos-etal05}, 
we select only those subhalos belonging to the hosts whose masses lie in the narrow range of $1\le M_{h}/(10^{14}\,h^{-1}\,M_{\odot})\le 3$. 

It has been found that the subhalos surrounded by the tides highly coherent along a eigenvector direction ($q_{i}\ge 0.9,\ ^{\forall} i\in\{1,2,3\}$) 
tend to have higher mean values of the virial-to-accretion mass ratios than their counterparts ($q_{i}< 0.9$), no matter what eigenvector direction is 
chosen. 
Our interpretation of this result is that the tides highly coherent along any eigenvector direction has an effect of obstructing the satellite infalls onto 
the hosts, which leads the satellites to experience the least severe mass-loss evolution in their post-infall stages. It has also been shown that the 
high tidal coherence along the third (second) eigenvector direction has the strongest (weakest) obstructing effect on the satellite infalls. 

The tides highly {\it incoherent} along a different eigenvector direction, however, has turned out to have a different effect. 
The tides highly {\it incoherent} along the first eigenvector direction ($q_{1}< 0.2$) have an effect opposite to the tides coherent along the same 
direction ($q_{1}\ge 0.9$) on the subhalo mass-loss evolution: the former facilitates the satellite infalls while the latter obstructs them, leading the 
subhalos surrounded by the former to lose much larger amount of masses after the infalls than those surrounded by the latter.  
In fact, the subhalos surrounded by the tides highly {\it incoherent} along the first eigenvector direction have been found to yield the lowest mean 
virial-to-accretion mass ratios.
Whereas, the tides highly {\it incoherent} along the third eigenvector direction ($q_{3}< 0.2$) have an effect of obstructing rather than facilitating 
the satellite infalls, similar to the tides highly coherent along the same direction ($q_{3}\ge 0.9$). 

It is shown that the simultaneous coherence or {\it incoherence} of the tides along two or three eigenvector directions have more complex 
effects on the subhalo mass-loss evolution. The high tidal coherence along the first eigenvector direction has been found to be synergic with the 
high tidal {\it incoherence} along the minor eigenvector direction ($q_{1}\ge 0.9$ and $q_{3}< 0.2$) in obstructing the satellite-infalls, yielding the highest 
mean virial-to-accretion mass ratios of the subhalos. Whereas, the high tidal {\it incoherence} along the first eigenvector direction has turned out to 
be discordant with both of the high tidal coherence and {\it incoherence} along the third eigenvector direction in facilitating the satellite infalls. 
The high tidal coherence along the second eigenvector direction have turned out to be synergic with the high tidal coherence along the 
first eigenvector direction in obstructing the satellite infalls, provided that the tides are not so coherent along the third eigenvector direction. 
Meanwhile, provided that the tides are not so coherent along the first eigenvector direction, the high tidal coherence along the second eigenvector 
direction has been found synergic with the high tidal coherence along the third eigenvector direction in facilitating the satellite infalls.  

Although the tides highly coherent along one of the three eigenvector direction have an obstructing effect on the satellite infalls,  the simultaneous 
coherence of the tides along all of the three eigenvector directions have been found not to reinforce the obstructing effect. The tides highly coherent 
along the first eigenvector direction and {\it incoherent} along the second and third eigenvector directions have been found more effective in 
obstructing the satellite infalls than the tides simultaneously coherent along all of the three eigenvector directions. 
The same is true for the simultaneous {\it incoherence} of the tides along all of the three eigenvector directions, which have been found not to 
reinforce the effect of facilitating the satellite infalls. The tides highly coherent along the third eigenvector direction and simultaneously {\it 
incoherent} along the first and second eigenvector direction have been found more effective in facilitating the satellite infalls than the tides 
simultaneously {\it incoherent} along all of the three eigenvector directions.

Determining the mean values of the local density contrasts, $\langle\delta\rangle$, and tidal anisotropies, $\langle e\rangle$, averaged over
the regions with different tidal coherences, we have found negligible differences in $\langle\delta\rangle$ and substantial differences in 
$\langle e\rangle$ among the regions. Noting that the simultaneous coherence along all of the three eigenvector directions plays a significant 
role of reducing the tidal anisotropy, and recalling that the high tidal anisotropy has been known to obstruct the satellite infalls \citep[e.g.,][]{zomg1}, 
we have explained that the higher tidal anisotropy should be contributed to the stronger obstructing effect of the tides highly coherent along 
the first eigenvector direction but highly {\it incoherent} along the second and third eigenvector directions than the tides highly coherent along all of 
the three eigenvector directions.  Yet, we have also shown that the multi-dimensional tidal coherence have an independent net effect on the 
subhalo-mass loss evolution, which cannot be ascribed simply to the differences in the tidal anisotropy. 

Given that the mean virial-to-accretion mass ratios of the subhalos reflect not only their mass-loss evolutions but also how fast their host clusters 
have grown as well as in what dynamical states they are \citep{bos-etal05}, the bottom line of our work is as follows: 
The formation and evolution of the cluster halos at fixed mass scales located in the environments with similar densities and tidal anisotropies 
still show variations with the multi-dimensional effects of the tidal coherence. We suspect that this result may be responsible for 
the large scatters around the spherical critical density contrast of $\delta_{c}\equiv 1.68$ required for the formation of a cluster halo, 
which could not be entirely explained by the scale-dependence of the non-spherical counter-part, $\delta_{ec}$
\citep[e.g.,][]{MR10,CA11}. 
Our result may be also closely related to the elusive nature of the large-scale assembly bias, whose existence have so far gained no 
observational confirmations \citep[e.g., see][]{SM19}. 
It is not only the density and tidal strengths but also the multi-dimensional tidal coherence that we must take into account to 
detect the large-scale assembly bias.
We plan to work on finding a direct link between the tidal coherence and the large-scale assembly bias as well as on 
extending the excursion set model by incorporating the tidal coherence, hoping to report the results elsewhere in the near future.

\acknowledgements

I acknowledge the support of the Basic Science Research Program through the National Research Foundation (NRF) 
of Korea funded by the Ministry of Education (NO. 2016R1D1A1A09918491).  I was also partially supported by a research 
grant from the NRF of Korea to the Center for Galaxy Evolution Research (No.2017R1A5A1070354).

\clearpage
\clearpage
\clearpage
\clearpage
\clearpage
\clearpage
\clearpage
\clearpage
\clearpage
\clearpage
\clearpage
\clearpage
\clearpage
\begin{deluxetable}{ccc}
\tablewidth{0pt}
\setlength{\tabcolsep}{5mm}
\tablecaption{1D Tidal Coherence, Mean Mass and Number of the Hosts.}
\tablehead{condition & $\langle M_{h}\rangle$ & $N_{h}$ \\
& $(10^{14}\,h^{-1}M_{\odot})$ & }
\startdata
$q_{1}\ge 0.9$ & $1.63\pm 0.03$ & $306$ \\
$q_{1}<0.9$ & $1.58\pm 0.02$ & $1096$\\
$q_{1}<0.2$ & $1.78\pm 0.01$ & $179$\\
\hline
$q_{2}\ge 0.9$ &$1.60\pm 0.04$ & $174$ \\
$q_{2}<0.9$ & $1.59\pm 0.01$ & $1228$\\
$q_{2}<0.2$ & $1.75\pm 0.01$ & $275$\\
\hline
$q_{3}\ge 0.9$ & $1.60\pm 0.03$ & $254$ \\
$q_{3}<0.9$ & $1.59\pm 0.02$ & $1148$\\
$q_{3}<0.2$ & $1.70\pm 0.01$ & $203$\\
\enddata
\label{tab:1d}
\end{deluxetable}
\clearpage
\begin{deluxetable}{ccc}
\tablewidth{0pt}
\setlength{\tabcolsep}{5mm}
\tablecaption{2D Tidal Coherence, Mean Mass and Number of the Hosts.}
\tablehead{condition & $\langle M_{h}\rangle$ & $N_{h}$ \\
& $(10^{14}\,h^{-1}M_{\odot})$ & }
\startdata
$q_{1}\ge 0.9,\ q_{2}\ge 0.9$ & $1.62\pm 0.06$ & $92$ \\
$q_{1}\ge 0.9,\ q_{2}<0.9$ & $1.63\pm 0.04$ & $214$\\
$q_{1}< 0.9,\ q_{2}\ge 0.9$ & $1.59\pm 0.04$ & $162$ \\
$q_{1}< 0.9,\ q_{2}<0.9$ & $1.58\pm 0.02$ & $934$\\
\hline
$q_{1}\ge 0.9,\ q_{2}<0.2$ & $1.58\pm 0.08$ & $26$\\
$q_{1}< 0.2,\ q_{2}\ge 0.9$ & $1.42\pm 0.19$ & $9$ \\
$q_{1}< 0.2,\ q_{2}<0.2$ & $1.63\pm 0.06$ & $78$\\
\enddata
\label{tab:2d}
\end{deluxetable}
\clearpage
\begin{deluxetable}{ccc}
\tablewidth{0pt}
\setlength{\tabcolsep}{5mm}
\tablecaption{3D Tidal Coherence, Mean Mass and Number of the Hosts.}
\tablehead{condition & $\langle M_{h}\rangle$ & $N_{h}$ \\
& $(10^{14}\,h^{-1}M_{\odot})$ & }
\startdata
$q_{1}\ge 0.9,\ q_{2}\ge 0.9,\ q_{3}\ge 0.9$ & $1.61\pm 0.06$ & $82$ \\
$q_{1}\ge 0.9,\ q_{2}\ge0.9,\ q_{3}< 0.9$ & $1.65\pm 0.17$ & $11$\\
$q_{1}\ge 0.9,\ q_{2}< 0.9,\ q_{3}\ge 0.9$ & $1.73\pm 0.19$ & $10$ \\
$q_{1}\ge 0.9,\ q_{2}<0.9,\ q_{3}< 0.9$ & $1.64\pm 0.04$ & $203$\\
$q_{1}< 0.9,\ q_{2}\ge 0.9,\ q_{3}\ge 0.9$ & $1.61\pm 0.16$ & $10$ \\
$q_{1}< 0.9,\ q_{2}\ge0.9,\ q_{3}< 0.9$ & $1.62\pm 0.06$ & $71$\\
$q_{1}< 0.9,\ q_{2}< 0.9,\ q_{3}\ge 0.9$ & $1.59\pm 0.04$ & $152$ \\
$q_{1}< 0.9,\ q_{2}<0.9,\ q_{3}< 0.9$ & $1.57\pm 0.02$ & $863$\\
\hline
$q_{1}\ge 0.9,\ q_{2}<0.2,\ q_{3}< 0.2$ & $1.54\pm 0.09$ & $22$\\
$q_{1}< 0.2,\ q_{2}\ge 0.9,\ q_{3}< 0.2$ & $1.42\pm 0.19$ & $9$\\
$q_{1}< 0.2,\ q_{2}< 0.2,\ q_{3}\ge 0.9$ & $1.58\pm 0.12$ & $19$ \\
$q_{1}< 0.2,\ q_{2}<0.2,\ q_{3}< 0.2$ & $1.61\pm 0.17$ & $5$\\
\enddata
\label{tab:3d}
\end{deluxetable}

\end{document}